# Plasmon-Driven Hot Electron Transfer at Atomically Sharp Metal-Semiconductor Nanojunctions

*Masiar Sistani[1], Maximilian G. Bartmann[1], Nicholas A. Güsken[2], Rupert F. Oulton[2],*

*Hamid Keshmiri[1], Minh Anh Luong[3], Zahra Sadre-Momtaz[4],*

*Martien I. den Hertog[4], Alois Lugstein[1]*

[1] Institute of Solid State Electronics, Technische Universität Wien, Gußhausstraße 25-25a, 1040 Vienna, Austria

[2] The Blackett Laboratory, Department of Physics, Imperial College London, London SW7 2AZ, United Kingdom

[3] Univ. Grenoble Alpes, CEA, INAC, MEM, F-38000 Grenoble, France

[4] Institut NEEL CNRS/UGA UPR2940, 25 avenue des Martyrs, 38042 Grenoble, France

E-Mail Address of corresponding author: alois.lugstein@tuwien.ac.at

# ABSTRACT

Recent advances in guiding and localizing light at the nanoscale exposed the enormous potential of ultra-scaled plasmonic devices. In this context, the decay of surface plasmons to hot carriers triggers a 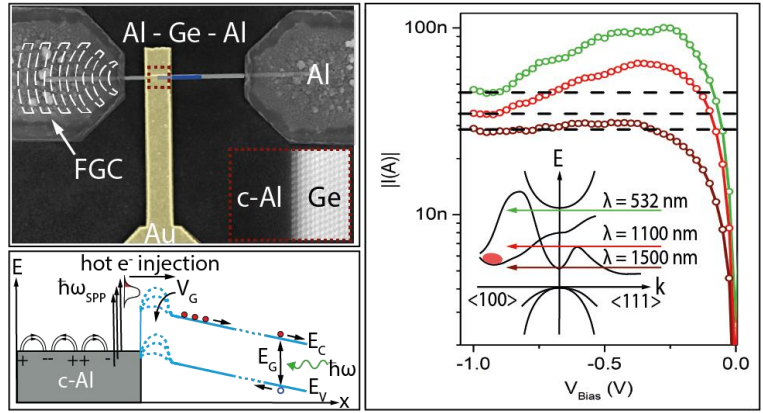 variety of applications in boosting the efficiency of energy-harvesting, photo-catalysis and photo-detection. However, a detailed understanding of plasmonic hot carrier generation and particularly the transfer at metal-semiconductor interfaces is still elusive. In this paper, we introduce a monolithic metal-semiconductor (Al-Ge) heterostructure device, providing a platform to examine surface plasmon decay and hot electron transfer at an atomically sharp Schottky nanojunction. The gated metal-semiconductor heterojunction device features electrostatic control of the Schottky barrier height at the Al-Ge interface, enabling hot electron filtering. The ability of momentum matching and to control the energy distribution of plasmon-driven hot electron injection is demonstrated by controlling the interband electron transfer in Ge leading to negative differential resistance.

Plasmonic-metal nanostructures have become a powerful tool to concentrate and manipulate light below the diffraction limit, paving a pathway to couple optical energy effectively to nanoscale systems.[1,2,3,4] The plasmonic manipulation of light in metal-semiconductor (M-S) heterostructures became of particular interest due to their capability of generating hot electrons in the metal that can be transferred to the semiconductor and facilitate chemical reactions.[5] The selection of momentum matched materials[6] and the design of optimized device architectures are thereby important for controlling plasmon-induced optical field distribution,[7] hot electron injection[8] and energy transfer at a M-S heterojunction.[9] Propagating surface plasmons approaching a M-S interface can decay either radiatively[10] via scattering of photons or non-radiatively through the generation of hot carriers.[11] For radiative decay, photons are scattered from the metal into the adjacent semiconductor driving electron-hole pair generation commonly denoted internal photogeneration.[12,13,14,15,16] However, this mechanism is limited to energies above the semiconductor band-gap[7] and for Ge due to the indirect bandgap the small electron-photon cross-section makes this process inherently inefficient.[17] For non-radiative plasmon decay, plasmon-driven electrons that simultaneously move toward the planar interface while also having sufficient kinetic energy to overcome the Schottky barrier, may be directly injected into the semiconductor.[17] These injected hot electrons exhibit a rather broad energy distribution, which stems from the almost continuous density of electronic energy states that exist below the Fermi level of metals.[15] However, assuming excitation of electrons at the Fermi level, the maximum energy of the injected hot electrons is $\mathbf{E} = \hbar\boldsymbol{\omega}_{\mathbf{SPP}}$,[7,17,18] with $\boldsymbol{\omega}_{\mathbf{SPP}}$ being the surface plasmon polariton (SPP) angular frequency.[8] Injection efficiencies less than 1% are expected due a range of limiting processes including, ultra-fast carrier-carrier relaxation,[5] the lack of vertical momentum of hot electrons and reflection at the M-S interface.[19] Despite a vast body of pioneering work on ab initio calculations[15,20] and experimental investigations[7,17,21,22,23] a complete understanding regarding plasmon decay is still elusive. Nevertheless, to make use of the full potential of surface plasmon-driven hot electron transport effects in applications such as photo-catalysis,[5] light harvesting[9] and photodetection[11] a detailed understanding of plasmon-driven hot electron generation is a prerequisite.[11]

RESULTS/DISCUSSION

In this work, we investigate surface plasmon-driven electron transfer processes at an atomically sharp M-S nanojunction of a monolithic Al-Ge nanowire (NW) heterostructure. Figure 1a and b show the schematic and a false-color scanning electron microscopy (SEM) image of the gated plasmon transfer device (GPTD) comprising a focused grating coupler (FGC) to launch SPPs in the single crystalline Al (c-Al) NW plasmon waveguide attached to the semiconducting Ge segment. An omega-shaped gate electrode directly atop of the atomically sharp heterojunction enables electrostatic control of the Schottky barrier at the M-S interface.[24] The device is realized on a 40 nm thick $Si_3N_4$ membrane[25] by a thermally induced exchange reaction between single-crystalline Ge NWs covered by a passivating $Al_2O_3$ shell and lithographically defined Al contact pads resulting in crystalline Al-Ge-Al NW heterostructures. More details regarding the Al-Ge heterostructure formation are described in the work of Kral et al.[26] and El Hajraoui et al.[27] The high angle annular dark field (HAADF) scanning transmission electron microscopy (STEM) images in Figure 1c-e show the monolithic Al-Ge-Al nanowire heterostructure with atomically sharp interfaces enwrapped by a 20 nm thick passivating $Al_2O_3$-shell.

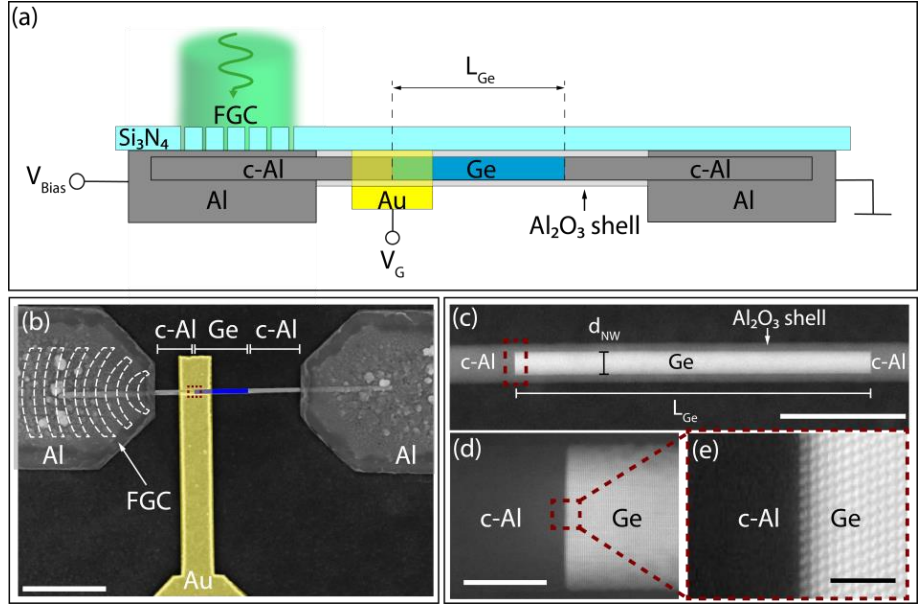

*Figure 1.* (a) Schematic illustration of the top-gated Al-Ge NW heterostructure device fabricated on a 40 nm thin $Si_3N_4$ membrane. (b) False color SEM image of the GPTD device. The FGC atop of the $Si_3N_4$ membrane is schematically indicated with dashed lines. Scale bar is 1 μm. (c) HAADF STEM image of the entire Al-Ge-Al NW heterostructure comprising a Ge segment length of $L_{Ge}$ = 600 nm and a diameter of $d_{NW}$ = 40 nm enwrapped by a 20 nm thick $Al_2O_3$ shell. Scale bar is 200 nm. TEM images showing (d) the entire Al-Ge interface of the NW heterostructure (scale bar is 40 nm) and a zoomed-in image where the Ge region is oriented on the [110] zone axis (scale bar is 5 nm).

Effective surface plasmon excitation is achieved by coupling normal incident TM polarized laser with a vacuum wavelength of λ = 532 nm into the c-Al NW waveguide using a $Si_3N_4$ membrane FGC,[28] located above the Al contact pad (see Figure 1a and 1b).[29] For such a configuration we experimentally determined a SPP propagation length of $L_m$=140 nm at λ = 532 nm for 40 nm thin c-Al NWs.[29]

Figure 2a shows a cross-sectional schematic of the Al plasmonic waveguide enwrapped by the Al$_2$O$_3$ shell and the omega-shaped metal gate. Finite Difference Time Domain (FDTD) simulations confirm a bound mode propagation along the NW, below the gate and right until the Al-Ge interface. Here, the entire propagation pathway of a TM polarized incident beam coupled via the FGC to the c-Al-NW and the gated Al-Ge NW junction was considered. Figure 2a shows the field intensity distribution of the coupled light in close vicinity to the gated Al-Ge interface where hot carriers are excited.

The omega-shaped top gate at the M-S interface enables tuning of the Schottky barrier height at the M-S heterojunction. In fact, the electrical characteristic of the heterojunction can be tuned between Ohmic and Schottky behaviors (supporting Figure S1). According to the schematic in Figure 2b, only hot electrons that move toward the nanojunction with sufficiently high energies overcome the barrier and enter the Ge segment, following nonradiative SPP decay (see schematic in Figure 2b). By applying a bias voltage (V$_{Bias}$), as shown in Figure 1a, for a given barrier height, these injected charge carriers induce a current, further denoted as SP-current. Thus with the GPTD we can detect surface plasmons by direct electrical means,[20,30] and even tune the energy of the electrons surpassing the Schottky junction of this M-S plasmon detector.

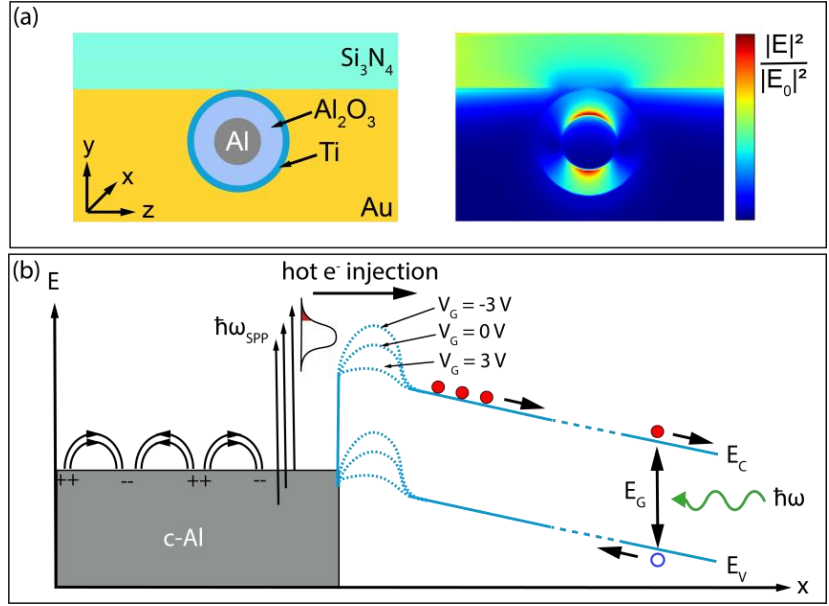

*Figure 2. (a) Illustration of the gated c-Al NW segment cross section in the vicinity of the Al-Ge nanojunction indicating the different material layers surrounding the NW. The right plot shows the simulated distribution of the normalized electric field intensity ($|E|^2/|E_0|^2$) at the Al-Ge interface for a TM fundamental mode coupled via the FGC ($\lambda$ = 532 nm). The simulations show that the mode is bound to the c-Al NW even below the gate. The scale bar is 40 nm. (b) Schematic illustration of SPP propagation along the c-Al NW, SPP decay induced carrier injection into the Ge segment and the band-diagram with gate dependent barrier height. The green arrow schematically shows photoexcitation of the Ge segment and the associated electron-hole pair generation.*

To demonstrate the mode of operation of the electrostatically tunable barrier, which works as an energetic filter for plasmon-driven hot electron injection, we performed wavelength dependent measurements. A tunable laser was focused on the FGC and the plasmon-driven hot electron transfer was investigated measuring the SP-current through the GPTD as a function of the gate potential. Figure 3 shows a comparison of the SP-current (red) with the photo-current (PH-current) induced by directly focusing the laser onto the Ge segment (blue dotted curves). A PH-current only occurs for photon energies higher than the bandgap and appeared to be independent of the gate voltage. A SP-current occurs for both interband excitation as well as plasmon-driven hot electron injection, which is tunable by the gate voltage from about 1500 nm to about 1850 nm. For increasing gate voltage, corresponding to a lowering of the Schottky barrier height, less energetic electrons can surpass the junction and contribute to the SP-current, even when the photon energy coupled into the c-Al waveguides is below the bandgap energy of Ge. This sub-bandgap photodetection rules out a mechanism based upon direct photogeneration at the M-S interface.

Assuming that the plasmon-induced near field drives the hot electron generation,[7,17] the maximum wavelength where SP-current can still be measured for a certain gate voltage, indicates both the maximum energy of injected electrons into the Ge as well as the Schottky barrier height. The inset in Figure 3 shows thus calculated effective barrier heights for plasmon-driven hot electron transfer from the c-Al waveguide into the Ge detector as a function of the gate voltage.

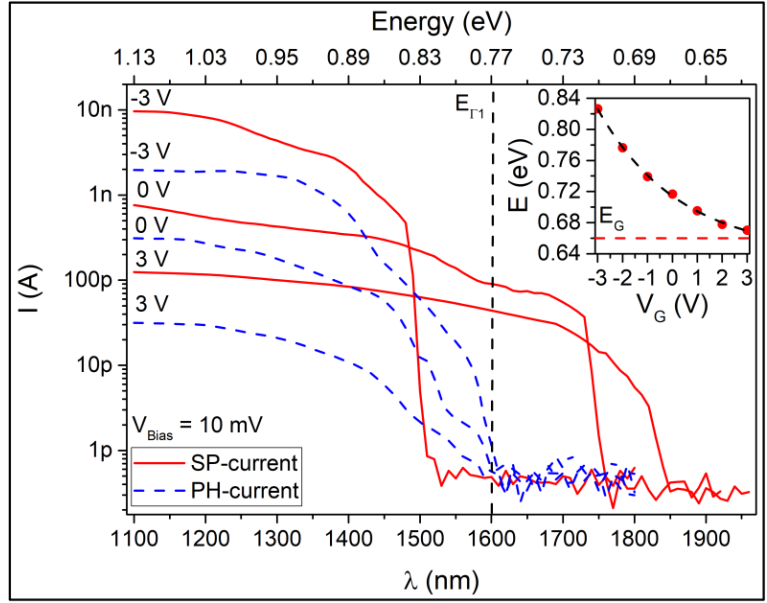

*Figure 3.* Spectral response of the SP- and PH-current for a bias voltage of $V_{Bias} = 10$ mV and gate-voltages between -3 V and 3 V. The inset shows the thereof evaluated energy of hot electrons extracted from wavelength dependent measurements of the SP-current.

The ability to control the energy and momentum of plasmon-driven hot electrons injected into the Ge enables us to induce negative differential resistance (NDR),[31] an effect that may be used for fast switching logic circuits, static memory cells, or high-frequency oscillators.[32] NDR can arise from electrons with sufficiently high energy being scattered to a heavy mass valley with a lower mobility, thus increasing the resistivity of the Ge.[33] This so-called transferred electron effect has been observed at high electric fields in numerous systems, perhaps best known as the Gunn-effect in GaAs[34] but also GaN nanocrystals[35] and Ni/Ge Schottky diodes.[36]

To investigate such SPP momentum induced NDR in Ge, the GPTD was operated as a Schottky barrier field effect transistor (FET). The black curve in the semi-logarithmic plot of Figure 4 shows the I/V characteristic with the drain current $I \propto \exp(eV/k_B T)$, which is typical for semiconductor NW devices with two back-to-back Schottky contacts.[37] The rectifying behavior is a direct consequence of different barrier heights of the M-S junctions interfacing the NW.[38,39,40] The SP-currents with laser illumination at the FGC at $\lambda = 532$ nm, $\lambda = 1100$ nm and $\lambda = 1500$ nm are shown in green, red and dark red respectively. For a gate voltage of 3 V and thus low Schottky barrier height, plasmon-driven hot electrons are effectively injected into the Ge segment, thus inducing a large SP-current. At positive bias voltages we observed a steady SP-current increase for all three wavelengths. The injected hot electrons arising from the plasmon decay at the Al-Ge interface are moving against the electric field and undergo thermalization via electron-electron and electron-phonon scattering as well as electron-hole pair generation (see lower inset in Figure 4).[41] These additional charge carriers are swept out from the Ge segment driven by the electric field and contribute to the overall current through the FET. For negative bias, NDR is observed, with a maximum peak to valley ratio of 2.3 for the SP-current recorded for plasmon excitation with a laser wavelength of $\lambda = 532$ nm (green curve). Note that for negative bias the plasmon-driven hot electrons are injected within momentum aligned to the electrical field in the Ge segment and are thus accelerated (upper inset).

According to the Ridley-Watkins-Hilsum theory[36], NDR via an electron transfer process requires the lower and upper valleys between which electrons are transferred to be separated by an energy difference much larger than the thermal energy but smaller than the energy gap of the semiconductor. This holds for Ge as the energy difference between the L-point and Δ-point minima of the ⟨111⟩ and ⟨100⟩ sub-bands of $\Delta E = 0.19$ eV[37] is significantly higher than the thermal energy at room temperature ($k_B T = 25.8$ meV) but much less than the bandgap ($E_G = 0.66$ eV) (inset of Figure 4b).[37] To finally achieve NDR, it is required that electrons in the lower valley have a lower effective mass than those in the upper valley.[34] For Ge the transverse effective electron mass in the Δ-point $m^*_{\Delta,t}$ = 0.288 $m_0$, is significantly higher than that in the L-point minimum, $m^*_{L,t}$ = 0.082 $m_0$.[42] Although the Γ-point minimum is energetically closer to the L-point minimum, the coupling constant between the 111 and 000 minima is significantly lower than between the 111 and 100 minima.[42] Thus, the transfer of electrons mainly occurs from the L-valley to the Δ-valley and the respective energy difference of ΔE = 0.19 eV[37] determines the threshold electric field for electron repopulation. Hot electrons arising from the plasmon decay at the Al-Ge interface are already at a higher energy level and require only moderate electric fields to overcome a smaller energy difference to be scattered to the Δ-valley. Accordingly, the NDR effect fades out as the energy of the hot electrons declines for exciting the FGC with laser light at λ = 1500 nm. Therefore, the maximum energy of injected hot electrons (0.82 eV) is not sufficient anymore to inject electrons effectively into the Δ-valley.

It is important to note that our measurements provide substantial experimental evidence that the momentum of hot electrons originating from surface plasmon decay is suitable for scattering into the lower mobility Δ-valley of Ge. In contrast, no signs of NDR could be found for directly focusing the laser onto the Ge segment (see supporting Figure S2).

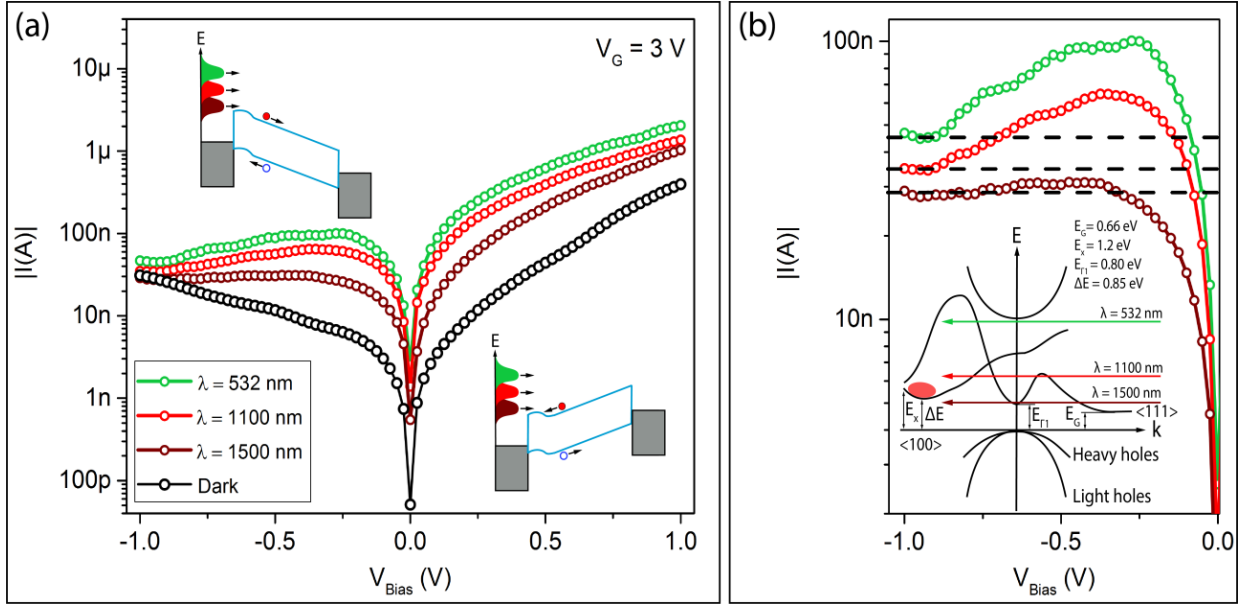

***Figure 4.*** (a) *Dark current and SP- current recorded for excitation wavelengths of λ = 532 nm, 1100 nm and 1500 nm at $V_G$ = 3 V. The insets show the respective band-diagrams for applying a positive and negative bias voltage to the heterostructure device. (b) Zoom showing the region of NDR. The inset shows the band diagram of Ge with the hot carrier injection into the ⟨100⟩ valley indicated.*

## CONCLUSIONS

In conclusion, we have demonstrated a near-field electrical SPP detector providing a platform to independently examine plasmon-induced hot carrier injection and photoexcitation. The system is based on a monolithic axial M-S-M heterostructure that features a precise control of the injection barrier height at an abrupt M-S interface, enabling the probing of the hot electron injection from SPP decay. This architecture allows an investigation of hot electron transport effects initiated by SPP decay at an atomically sharp M-S heterojunction. By precisely controlling the injection barrier and SPP excitation, we demonstrated SPP momentum induced NDR. In contrast, no signs of NDR could be found for directly focusing the laser onto the Ge segment. Thus, the demonstrated investigations provide a route to a better understanding of plasmonic hot electron devices and may pave the way for novel device concepts based on a plasmon-induced NDR.

METHODS/EXPERIMENTAL

Device fabrication:

The starting materials were VLS grown Ge NWs with diameters between 30-50 nm coated with 20 nm high-k $Al_2O_3$ using atomic layer deposition. The Ge NWs were drop casted onto 40 nm thin $Si_3N_4$ membranes[25] and contacted by Al pads fabricated by electron beam lithography, 100 nm Al sputter deposition and lift-off techniques. A successive thermally induced exchange reaction by rapid thermal annealing at a temperature of T = 624 K in forming gas atmosphere initiates the substitution of the Ge core by c-Al.[26,27] An omega shaped Ti/Au top-gate was fabricated above the M-S interface using electron beam lithography, Ti/Au electron-beam evaporation (8 nm Ti, 100 nm Au) and lift-off techniques. To complete the device, the FIB column of a Zeiss Neon 40EsB CrossBeam system was used to pattern the FGC in the $Si_3N_4$ membranes. The milling current was set to 50 pA.

High-resolution HAADF STEM:

HAADF STEM was performed on Al-Ge-Al NW heterostructures fabricated on 40 nm thick $Si_3N_4$ membranes[25] using a probe corrected FEI Titan Themis, working at 200 kV. The Al-Ge interface in the shown images was oriented along the [110] direction of observation of the Ge crystal.

Electrical and optical characterization:

The biasing of the proposed Al-Ge-Al NW heterostructures was performed using a Keysight B1500A semiconductor analyzer. For optical excitation, a frequency doubled Nd:YAG laser emitting linearly polarized light at $\lambda = 532$ nm was coupled into a WITec Alpha300 and focused on the device through a Zeiss 100x objective (NA = 0.75, WD = 4 mm) enabling a diffraction limited spot size of ~865 nm. The output power of the system was set to have negligible laser heating effects in the NW. For investigating the spectral response of the plasmon current, the white light from a broadband laser source (SuperK Extreme, NKT) fiber was coupled to a monochromator (SuperK Select, NKT). The output of this system is supplied to a NKT SuperK Select acoustic-optical tunable filter (AOTF). The system consists of three AOTFs, which act as monochromators with separated channels for visible ($\lambda = 500$ nm - 700 nm), near-infrared ($\lambda = 600$ nm - 1100 nm) and infrared ($\lambda = 1100$ nm - 2000 nm) light. The output from the ATOFs is coupled into a WITec Alpha300. The beam is passing a 50-50 beam splitter and is focused on the sample through a Zeiss 100x objective (NA = 0.75, WD = 4 mm). The output is focused on the device with power densities chosen to have negligible laser heating effects in the NW.

FDTD simulations:

Numerical 3D FDTD calculations were performed using commercial software (Lumerical). The entire 3D system including the incoupling FGC and the Al-Ge-Al NW heterostructure were considered. A Gaussian beam served as source with a central wavelength of $\lambda = 532$ nm, illuminating the FGC. The mesh size was set to 0.4 nm, allowing to account for the minimum feature sizes of the structure. The electric field intensities indicated are normalized by the source intensity value $|E_0|^2$.

ASSOCIATED CONTENT

Supporting Information Available:

I/V measurements showing the ability to tune the Ge detector from an ohmic behavior to a Schottky behavior, PH-current measurement at $V_G = 3$ V. This material is available free of charge via the Internet at http://pubs.acs.org.

AUTHOR INFORMATION

**Corresponding Author**

*E-mail: alois.lugstein@tuwien.ac.at

**Author Contributions**

M.S. and M.G.B. fabricated the devices and conducted the measurements. N.A.G. conducted the FDTD simulations of the electric field intensity of the field distributions of the fundamental mode of the NW-$Si_3N_4$ membrane system. N.A.G. and R.F.O. contributed to the explanation of the underlying physical mechanisms. H.K. designed the FGC. Z.S.M. fabricated the $Si_3N_4$ membranes and M.A.L. and M.I.H. carried out TEM characterization. A.L. conceived the project, contributed essentially to the experimental design and provided expertise on theoretical interpretations. The manuscript was written through contributions of all authors. All authors have given approval to the final version of the manuscript.

**Notes**

The authors declare no competing financial interest.

ACKNOWLEDGMENT

The authors gratefully acknowledge financial support by the Austrian Science Fund (FWF): project No.: P28175-N27. The authors further thank the Center for Micro- and Nanostructures for providing the cleanroom facilities. We acknowledge support from the Laboratoire d'excellence LANEF in Grenoble (ANR-10-LABX-51-01). Financial support from the ANR-COSMOS (ANR-12-JS10-0002) project is acknowledged. We benefitted from the access to the Nano characterization platform (PFNC) in CEA Minatec Grenoble in collaboration with the LEMMA/IRIG group. We acknowledge support from Campus France in the framework of PHC AMADEUS 2016 for PROJET N° 35592PB. This project has received funding from the European

Research Council (ERC) under the European Union's Horizon 2020 research and innovation program (grant agreement N° 758385) for the e-See project.